\documentclass[preprint]{aastex631}
\usepackage[section]{placeins}
\usepackage{CJK}
\begin{document}
\begin{CJK*}{UTF8}{gbsn}
\title{Why ``solar tsunamis" rarely leave their imprints in the chromosphere}

\correspondingauthor{Ruisheng Zheng}
\email{ruishengzheng@sdu.edu.cn}

\author[0000-0002-2734-8969]{Ruisheng Zheng}
\affiliation{Institute of Space Sciences, Shandong University, Weihai 264209, China}
\affiliation{Institute of Frontier and Interdisciplinary Science, Shandong University, Qingdao 266237, China}
\affiliation{CAS Key Laboratory of Solar Activity, National Astronomical Observatories, Chinese Academy of Sciences, Beijing 100101, China}

\author[0000-0001-7887-5024]{Yihan Liu}
\affiliation{Institute of Space Sciences, Shandong University, Weihai 264209, China}

\author{Wenlong Liu}
\affiliation{Institute of Space Sciences, Shandong University, Weihai 264209, China}

\author{Bing Wang}
\affiliation{Institute of Space Sciences, Shandong University, Weihai 264209, China}

\author[0000-0003-4804-5673]{Zhenyong Hou}
\affiliation{School of Earth and Space Sciences, Peking University, Beijing 100871, China}

\author[0000-0002-9634-5139]{Shiwei Feng}
\affiliation{Institute of Space Sciences, Shandong University, Weihai 264209, China}

\author[0000-0003-1034-5857]{Xiangliang Kong}
\affiliation{Institute of Space Sciences, Shandong University, Weihai 264209, China}
\affiliation{Institute of Frontier and Interdisciplinary Science, Shandong University, Qingdao 266237, China}

\author[0000-0002-2358-5377]{Zhenghua Huang}
\affiliation{Institute of Space Sciences, Shandong University, Weihai 264209, China}

\author[0000-0001-5705-661X]{Hongqiang Song}
\affiliation{Institute of Space Sciences, Shandong University, Weihai 264209, China}
\affiliation{Institute of Frontier and Interdisciplinary Science, Shandong University, Qingdao 266237, China}

\author[0000-0002-1369-1758]{Hui Tian}
\affiliation{CAS Key Laboratory of Solar Activity, National Astronomical Observatories, Chinese Academy of Sciences, Beijing 100101, China}
\affiliation{School of Earth and Space Sciences, Peking University, Beijing 100871, China}
\affiliation{State Key Laboratory of Space Weather, National Space Science Center, Chinese Academy of Sciences, Beijing 100190, China}

\author[0000-0002-7289-642X]{Pengfei Chen}
\affiliation{School of Astronomy and Space Science, Nanjing University, Nanjing 210023, China}
\affiliation{Key Laboratory of Modern Astronomy and Astrophysics (Nanjing University), Ministry of Education, Nanjing 210023, China}

\author[0000-0003-3439-4127]{Robertus Erd\'{e}lyi}
\affiliation{Solar Physics \& Space Plasma Research Center (SP2RC), School of Mathematics and Statistics, University of Sheffield, Hicks Building, Hounsfield Road, S3 7RH, United Kingdom}

\author[0000-0001-6449-8838]{Yao Chen}
\affiliation{Institute of Space Sciences, Shandong University, Weihai 264209, China}
\affiliation{Institute of Frontier and Interdisciplinary Science, Shandong University, Qingdao 266237, China}



\begin{abstract}

Solar coronal waves frequently appear as bright disturbances that propagate globally from the eruption center in the solar atmosphere, just like the tsunamis in the ocean on Earth. Theoretically, coronal waves can sweep over the underlying chromosphere and leave an imprint in the form of Moreton wave, due to the enhanced pressure beneath their coronal wavefront. Despite the frequent observations of coronal waves, their counterparts in the chromosphere are rarely detected. Why the chromosphere rarely bears the imprints of solar tsunamis remained a mystery since their discovery three decades ago. To resolve this question, all coronal waves and associated Moreton waves in the last decade have been initially surveyed, though the detection of Moreton waves could be hampered by utilising the low-quality H$\alpha$ data from Global Oscillations Network Group. Here, we present 8 cases (including 5 in Appendix) of the coexistence of coronal and Moreton waves in inclined eruptions where it is argued that the extreme inclination is key to providing an answer to address the question. For all these events, the lowest part of the coronal wavefront near the solar surface appears very bright, and the simultaneous disturbances in the solar transition region and the chromosphere predominantly occur beneath the bright segment. Therefore, evidenced by observations, we propose a scenario for the excitation mechanism of the coronal-Moreton waves in highly inclined eruptions, in which the lowest part of a coronal wave can effectively disturb the chromosphere even for a weak (e.g., B-class) solar flare.

\end{abstract}



\section{Introduction} \label{sec:intro}
The solar atmosphere is composed of magnetized plasma and hosts a variety of waves that are generally believed to account for many fundamental processes in solar physics~\citep{DePontieu2007, Jess2009}. One of the most spectacular waves are the solar coronal waves often detected as circular disturbances traveling outward from their eruption source~\citep{Moses1997, Thompson1998}, similar to tsunamis on the Earth. Coronal waves are often associated with large-scale coronal mass ejections (CMEs), flares, or even small-scale jets. They have been interpreted differently, from e.g. the true fast-mode magnetohydrodynamic (MHD) waves~\citep{Wu2001, Ofman2002}, pseudo waves~\citep{Attrill2007, Delannee2008}, or hybrid waves~\citep{Chen2011, Liu2014, Warmuth2015}. Fast-mode coronal waves, by means of solar magneto-seismology, can yield estimates of the coronal magnetic field strengths that are hard to measure directly~\citep{Nakariakov1999, Ballai2007}, and may also accelerate solar energetic particles that influence space weather~\citep{Carley2013}.

Before their observational discovery, the existence of coronal waves was already conjectured by the detection of the chromospheric Moreton waves, first detected in the 1960s~\citep{Moreton1960, MoretonRamsey1960, Meyer1968}. A typical Moreton wave forms not far ($\sim$100 Mm) from an eruption center, and appears as an arc-shaped chromospheric disturbance. They propagate with a high speed ($\sim$1000 km s$^{-1}$) for minutes before becoming irregular and diffusive. Moreton waves are interpreted as the ground tracks of coronal fast-mode MHD wavefronts sweeping over the chromosphere~\citep{Uchida1968, Uchida1974}. In theory, the coronal wavefront leads to local enhanced pressure that compresses the chromospheric plasma, and therefore these Moreton-wave fronts appear in emission in the line core and blue wing of H$\alpha$, and in absorption in the red wing~\citep{Vrsnak2002, Vrsnak2008}. In addition, Moreton waves are frequently associated with solar type II radio bursts, which indicates that their coronal counterparts have evolved into shocks~\citep{Mann1995}.

With the high-quality observations from the Ahead (-A) and Behind (-B) spacecraft of Solar Terrestrial Relations Observatory~\citep{Kaiser2008} (STEREO) and the Solar Dynamics Observatory~\citep{Pesnell2012} (SDO), hundreds of coronal waves have been found and studied. However, only tens of Moreton waves have been observed. Since chromospheric Moreton waves are the counterparts of coronal waves, some fundamental questions rise naturally: Why did most coronal waves fail to ignite a detectable chromospheric disturbance? Why was no circular-shaped Moreton wave observed for all the coronal waves with circular-shaped fronts? Moreover, why are all previous Moreton waves only associated with strong M- and X-class flares? On the other hand, if a coronal wave can compress the chromosphere, the transition region (TR) as the interface between them should be compressed first, but the wave signatures in TR are also rarely reported. Hence, although it is generally believed that Moreton waves are the chromospheric counterpart of coronal waves, the bridge between two phenomena still remains elusive.

Combining observations from the dual perspectives by SDO and STEREO, we initially surveyed all coronal waves observed in the last decade from 2010 to 2021 (\url{https://www.lmsal.com/nitta/movies/AIA\_Waves/oindex.html}), and the associated Moreton waves were checked by the line core of H$\alpha$ from Global Oscillations Network Group (GONG). Note that most of the detected Moreton waves are only distinguished in animations and very faint in still images, due to the low-quality data of GONG. Though the low-quality data could decrease the detection number of total Moreton waves, the continuous observations can be related to all the coronal waves. { Because the same condition of low-quality GONG data are used to detect the related Moreton waves for all coronal waves that are studied here, the data factor can be neglected when we analyse the relationship between coronal waves and Moreton waves.}  The fraction of coronal waves associated Moreton waves is in progress in the upcoming statistics work.

In this Letter, we picked three cases of distinct chromospheric wave signatures and coronal waves associated with inclined eruptions that are identified by the observations in dual perspectives from the sample in the quick view, to first propose a scenario of the plausible excitation mechanism for Moreton waves. In this scenario, Moreton waves are easily ignited by the enhanced pressure behind coronal waves when the eruptions are highly ``inclined" and thus the nose parts of coronal waves are close to the solar surface. The strong reliance to highly inclined eruptions can reasonably explain the rarity of detectable Moreton waves.

\section{Results} \label{sec:results}

Three coronal wave events, presented and analyzed here in details, are respectively related to an X6.9-class flare on 2011 August 9 (E1), an M1.4-class flare on 2021 May 22 (E2), and a B9.5-class flare on 2011 July 3 (E3). Each event is accompanied by a CME and a type II radio burst (Supplementary Fig.~\ref{fs1} and Fig.~\ref{fs2}). The inclined eruptions and coronal waves are shown clearly in dual perspectives from Atmospheric Imaging Assembly (AIA)~\citep{Lemen2012} on-board SDO and from Extreme Ultraviolet Imager (EUVI)~\citep{Howard2008} on STEREO-A (Fig.~\ref{f1} and Fig.~\ref{f1} animation). In the full-disk composite images in AIA 304, 171, and 193~{\AA} (left panels), all the eruptions move highly nonradially (the sectors), and the inclined angles with respect to the solar radial direction are in the range of $\sim$63--76$^{\circ}$ estimated by the reconstruction of the eruption cores or lowest part of coronal wavefronts (Supplementary Fig.~\ref{fs3} and Fig.~\ref{fs4}). Following the inclined eruptions, the coronal waves also propagate unilaterally. In the local difference images of AIA 193~{\AA} and EUVI-A 195~{\AA} (middle and right columns in Fig.~\ref{f1}), the wavefronts are arc-like (white arrows), and their lowest parts close to the solar surface became very bright (red arrows).

Interestingly, different from the rest of the wavefront, the bright segments in AIA 193~{\AA} had evident associated signatures in AIA 171, 304~{\AA} and H$\alpha$ (Fig.~\ref{f2} and Fig.~\ref{f2} animation). In AIA 171, the wavefronts are replaced by coronal dimmings, which can be interpreted by the heating of local plasma~\citep{Liu2014}. The wave signatures in AIA 304~{\AA} are evident, and the H$\alpha$ responses are faint. The signatures in He II 304~{\AA} and H$\alpha$ can respectively represent the TR and the chromosphere counterparts of coronal waves, and both have arc-shapes similar to the bright segments of 193~{\AA} wave, indicating an intimate relation between three waves in different layers. They have similar speeds in the range of $\sim$380--580 km s$^{-1}$ according to the time-distance plots (Supplementary Fig.~\ref{fs5}). Assuming a fast magnetosonic speed of $\sim$200--300 km s$^{-1}$ at a low height for the bright segments~\citep{Wu2001}, the coronal waves may manifest as shocks with a fast-magnetosonic Mach number of $\sim$2~\citep{Warmuth2001}, in line with the occurrence of the radio type II bursts in Supplementary Fig.~\ref{fs2}. Note that the disturbance in E3 is the first example of a Moreton wave associated with a weak B-class flare, though the wave signal in H$\alpha$ is nearly invisible in the time-distance plot (Supplementary Fig.~\ref{fs5}).

For the coronal wavefronts (cyan boxes in Fig.~\ref{f2}), we analyse the evolution of intensity and emission at different passbands (Fig.~\ref{f3}). In the upper row, the curves represent the changes of the normalized intensities in AIA 193 (green), 171 (black), 304~{\AA} (red) , and GONG H$\alpha$ (purple). After the wave arrival (pink solid lines), the intensity of AIA 171~{\AA} begins to decrease suddenly, and the intensities in AIA 193 and 304~{\AA} increase abruptly. Though the rise onset of the intensity in GONG H$\alpha$ is uncertain due to the poor data quality, each of them achieves a distinct peak as those in AIA 193 and 304~{\AA} (dashed lines). The peaks in AIA 193, 304~{\AA}, and GONG H$\alpha$ are almost simultaneous for E1 and E2. For E3, the peak time in AIA 304~{\AA} is 36 seconds later than that in AIA 193~{\AA}, but 30 seconds earlier than that in GONG H$\alpha$ (panel (c)).

The bottom row of Fig.~\ref{f3} shows the average evolution of the differential emission measure (DEM) in the region of bright segments (cyan boxes in Fig.~\ref{f2}) obtained by employing the sparse inversion code~\citep{Cheung2015, Su2018}. After the wave passage (pink solid lines), the DEM at $\log(T/K) \approx 6.2$ increases obviously, and the DEM at $\log(T/K) \approx 6.0$ decreases simultaneously (white and pink arrows). It indicates that the local plasma of wavefronts is heated from $\log(T/K) \approx 6.0$ to $\log(T/K) \approx 6.2$, consistent with the bright wavefront in AIA 193~{\AA} and the dimmings in AIA 171~{\AA}. Intriguingly, the DEM at $\log(T/K) \approx 5.7$ also increases simultaneously with that at $\log(T/K) \approx 6.2$ (red arrows), though the increase in E3 is weak. Moreover, the continuous increase of DEM at $\log(T/K) \approx 5.7$ seemed to be evolved from that at {\bf $\log(T/K) \approx 5.9$}, which possibly implies their compression relationship. It indicates that the DEM increase at $\log(T/K) \approx 5.7$  likely represents the emission enhancement for the plasma at the bottom of the corona and/or the upper TR, due to the downward expansion and compression of the bright segment of the coronal wave that is nearly horizontal and close to the solar surface.


\section{Discussion and Conclusions}
Vr{\v s}nak et al.~\citep{Vrsnak2016} conjectured that maybe an asymmetric/nonradial eruption is an additional requirement for the generation of the Moreton wave in a weak eruption, when they simulated the Moreton waves associated with impulsive eruptions. In the few cases of coronal and Moreton waves observed by SDO and ground-based observatories, all the related eruptions are non-radial~\citep{Asai2012, Shen2012, Francile2016, Cabezas2019, Long2019, Wang2020}, and the inclined eruptions could be the result of filaments erupting into weaker magnetic field regions~\citep{Panasenco2013}. For instance, the inclined eruption and associated Moreton wave on 2014 March 29 is showed in Supplementary Fig.~\ref{fs6} and Fig.~\ref{fs6} animation), which has been studied by other researchers~\citep{Francile2016, Cabezas2019, Long2019}.

\begin{figure}[!ht]
\centering
\includegraphics{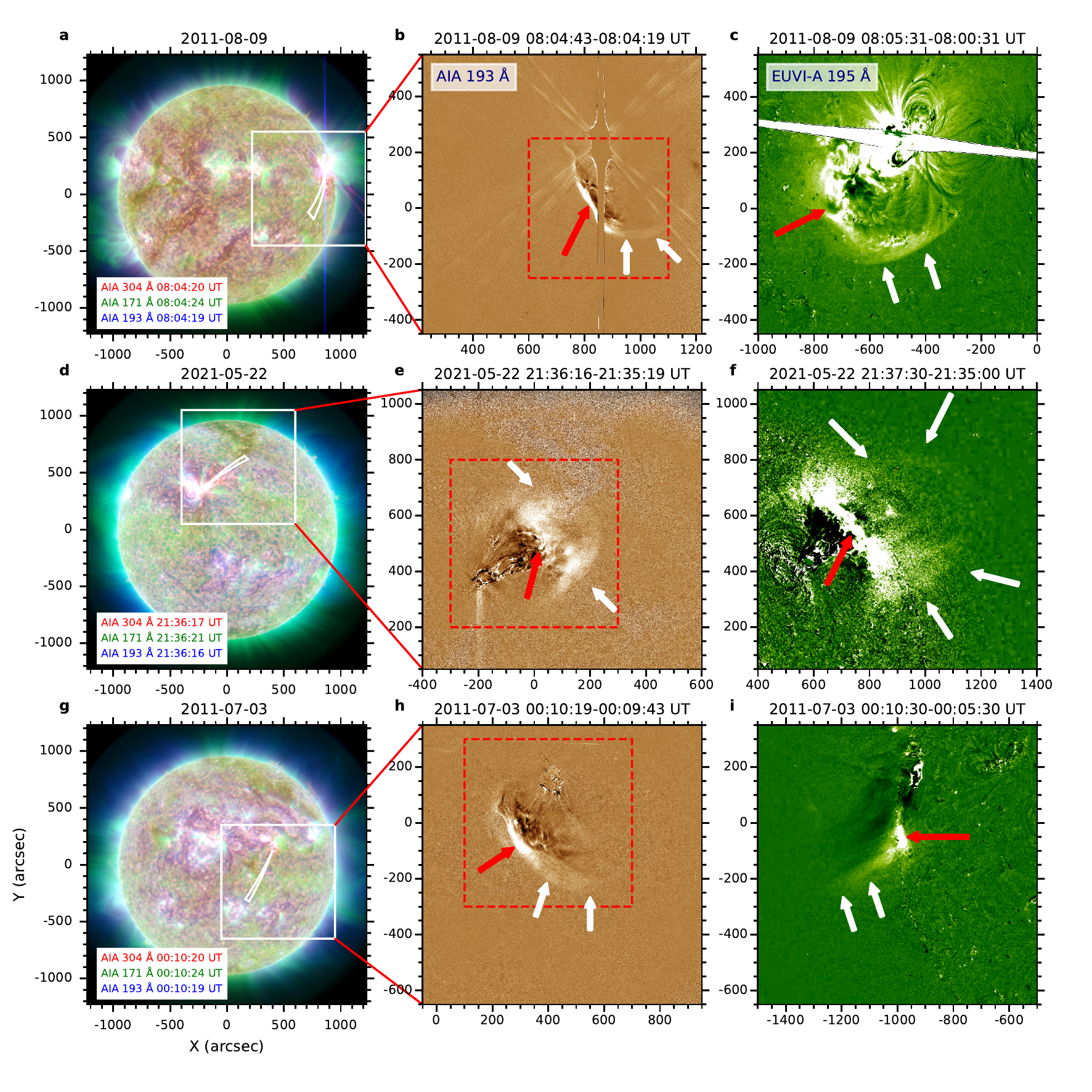}
\caption{{\bf Coronal waves in dual perspectives}. Left column: The eruptions are displayed in the full-disk composite images of AIA 304 (red), 171 (green), and 193~{\AA} (blue). The sectors indicate the directions of eruptions and associated coronal waves, and the white boxes represent the field of view (FOV) of middle panels. Middle and right columns: Coronal waves (white arrows) and bright segments (red arrows) are shown in difference images in AIA 193~{\AA} and EUVI-A 195~{\AA}. The red dashed boxes indicate the FOV of panels in Fig.~\ref{f2}. (An animation of middle and right columns is available online.)
}\label{f1}
\end{figure}

\begin{figure}[!ht]
\centering
\includegraphics{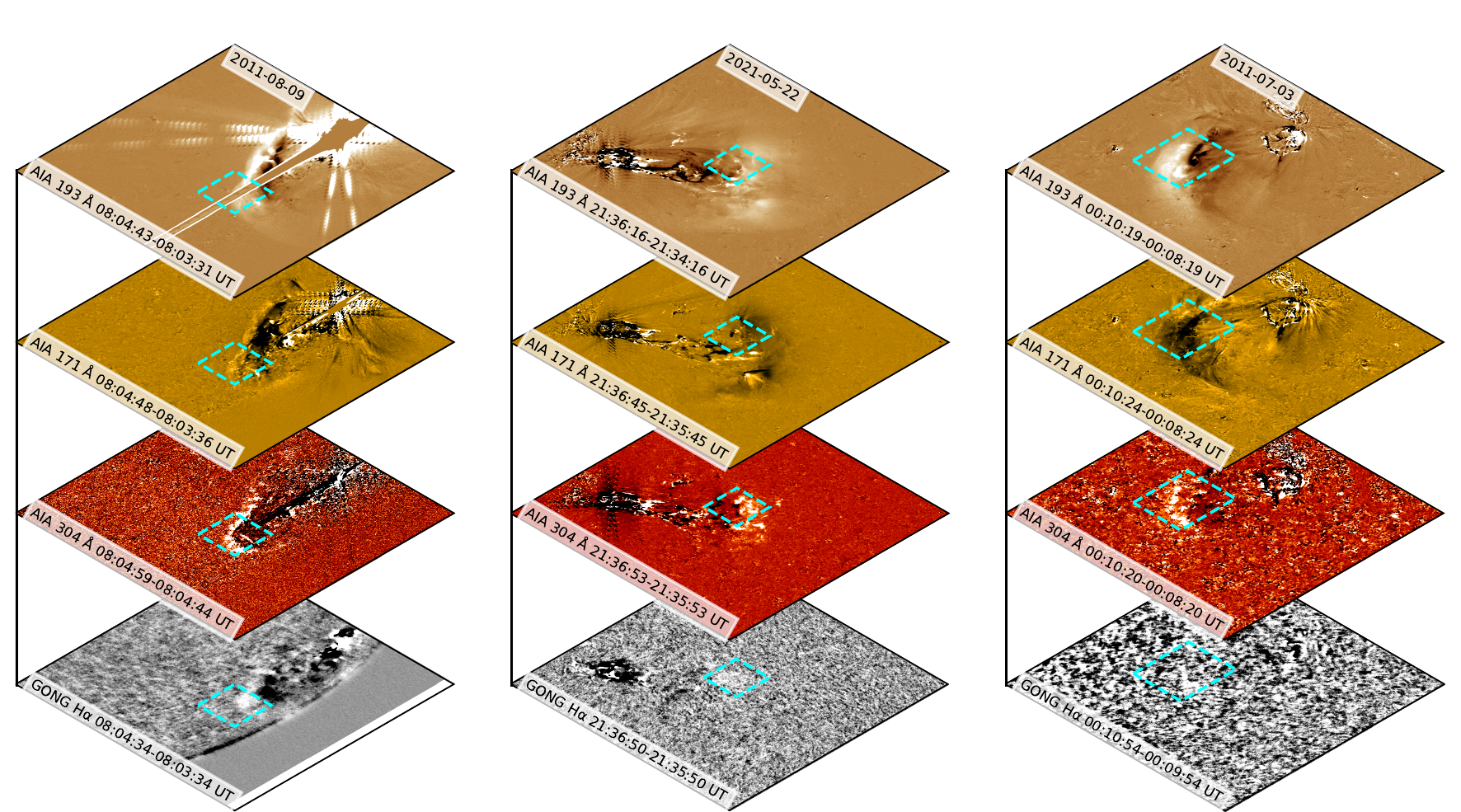}
\caption{{\bf The wave signatures in different atmospheric layers.} The cyan boxes in difference images outline the bright segment of coronal waves in AIA 193~{\AA} and their responses in AIA 171, 304~{\AA} and GONG H$\alpha$. The cyan boxes indicate the regions for the DEM evolution in Fig.~\ref{f3}. (An animation of associated with this figure is available online.)
}\label{f2}
\end{figure}

The AIA 304~{\AA} channel is mainly dominated by the two He II 303.8~{\AA} lines with a temperature of $\log(T/K) \approx 4.7$ for coronal holes, quiet Sun regions, active regions, and flare plasma. However, in QS off-limb the channel can also make an important contribution by Si XI 303.33~{\AA} line at the coronal temperature of $\log(T/K) \approx 6.2$ \citep{O'Dwyer2010}. Some previous observations have shown that wave signatures in AIA 304~{\AA} channel could be dominated in a coronal line of Si XI 303.33~{\AA} \citep{Long2010}, or the He II 303.8~{\AA} lines in the transition region \citep{Long2011}. In this Letter, it is very clear that the wavefronts in 304~{\AA} were { narrow}, similar to wavefronts in H$\alpha$, and were different with {wavefronts} in 193~{\AA}. Like Moreton waves, the wavefronts in 304~{\AA} were narrow, and closely related to the bright parts of coronal wavefronts (red and white arrows in Fig.~\ref{f1} and the dashed boxes in Fig.~\ref{f2}, similar to the case in \cite{Hou2022}. In addition, the wavefronts in 304~{\AA} can be shown in the disk in the perspective of SDO or STEREO. Hence, the morphological difference in the disk between the wavefronts in 304 and 193~{\AA} suggested that the wavefronts in 304~{\AA} in this Letter were much likely dominated by the He II line.

To further uncover the relationship between the coronal and Moreton waves, we check four more coronal waves that are associated with either inclined limb eruptions or radial eruptions. In the inclined limb eruptions (Supplementary Fig.~\ref{fs7} and Fig.~\ref{fs7} animation), the bright segments of coronal waves and disturbances in TR are prominent (blue arrows), though the chromospheric wave signatures are hard to identify due to the poor H$\alpha$ data quality. Although the radial eruptions are associated with strong X- or M-class flares (Supplementary Fig.~\ref{fs8} and Fig.~\ref{fs8} animation), no wave signature is found in He II 304~{\AA} and H$\alpha$. This is then consistent with the fact that the bulk of coronal waves is obvious in AIA 193 and 211~{\AA}, while there is no signature in AIA 304~{\AA}. According to Uchida's model~\citep{Uchida1968, Uchida1974}, the TR should be compressed first if the coronal wave can disturb the chromosphere. Therefore, the waves in He II 304~{\AA} for the above cases likely represent the response in TR, and can be regarded as the missing link between coronal and Moreton waves. Furthermore, for the above cases of inclined eruptions, all coronal waves successfully disturb the TR and chromosphere, including the one associated with the weak B-class flare. This further implies that the inclined eruption is the key to generating chromospheric Moreton waves.

\begin{figure}[!ht]
\centering
\includegraphics{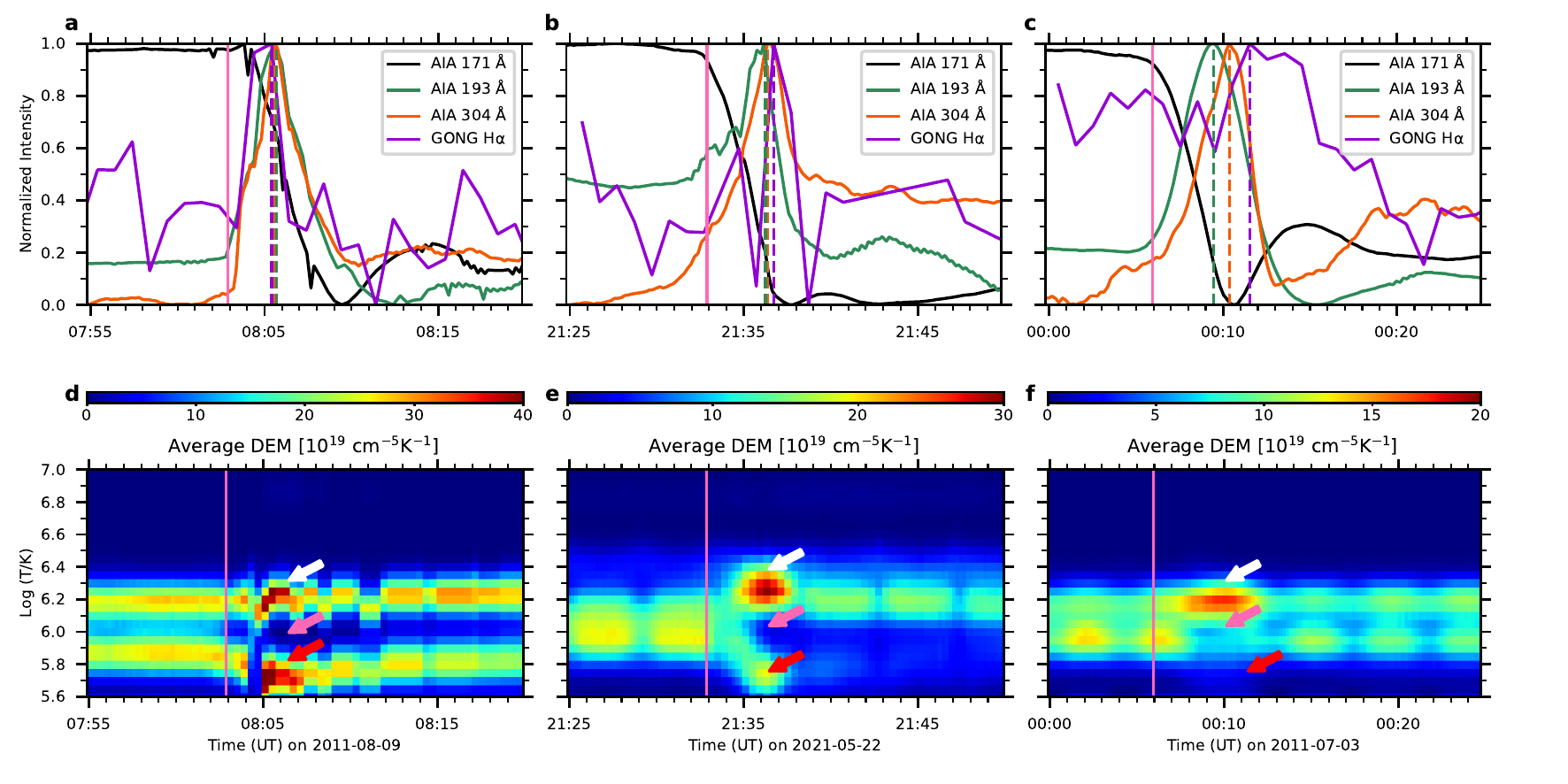}
\caption{{\bf The evolution for the intensity and emission of plasma.} Upper row: The light curves in the region of bright segments (cyan boxes in Fig.~\ref{f2}) in AIA 193, 171, 304~{\AA} and GONG H$\alpha$. Bottom row: The average DEM evolutions in the region of bright segments (cyan boxes in Fig.~\ref{f2}).
The pink solid lines indicate the arrival times of coronal waves, and the colored dashed lines represent the peak times in the associated passbands, and the colored arrows show the DEM changes at different temperatures.
}
\label{f3}
\end{figure}

For coronal waves in the inclined eruptions presented above, they display a bright segment of wavefronts near the solar surface, distinct from other parts (Fig.~\ref{f1}), and the disturbances in the TR and chromosphere simultaneously occur just under the brightest segment (Fig.~\ref{f2}). Comparing with the evolution in different regions of wavefronts, it is likely that the average DEM evolution in the region of brightest segment reveals the emission enhancement in the coronal base and the upper TR behind the coronal shocks/waves (Fig.~\ref{f3}), consistent with the scenario of the Moreton wave~\citep{Vrsnak2002}. Hence, the bright segments of coronal wavefronts and the related continuous increase of DEM indicate that the TR and chromosphere can be effectively disturbed by the lowest part of the inclined coronal waves, and provide further evidence of the role of inclined eruptions in driving the Moreton waves.

\begin{figure}[!ht]
\centering
\includegraphics{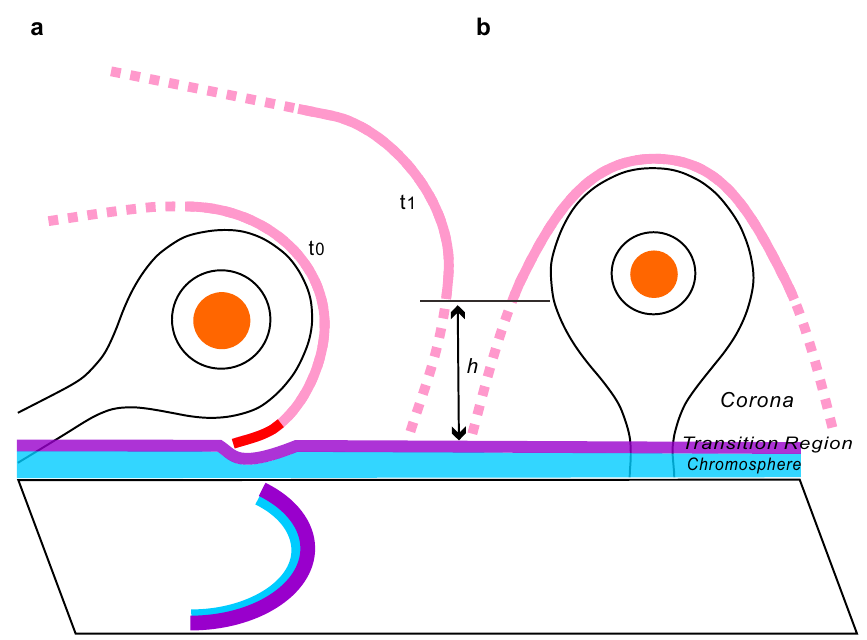}
\caption{{\bf The scenario of exciting mechanism for Moreton waves.} {\bf a)}: For an inclined eruption with a core (orange), the lowest part of the associated coronal wave (the solid-dotted pink front) becomes very bright (the red patch) at the onset (t0), and compresses the TR (the purple layer) and chromosphere (the blue layer) below. The projected purple and blue arcs represent the disturbances in the TR and chromosphere. The nose part of the coronal wavefront reaches a height of $h$ from the coronal base after the propagation in a few minutes (t1). {\bf b)}: For a radial eruption, the nose of the coronal wave forms at an initial height of $h$. Note that the coronal wave degenerates from a shock wave near the nose part (solid line) to an ordinary wave further away (dotted line).
}
\label{f4}
\end{figure}

Based on the above observational results, we propose a plausible excitation mechanism for Moreton waves in Fig.~\ref{f4}. The coronal wave (the solid-dotted pink curve) consists of a shocked nose (solid part) and weak flanks (dotted parts). For the coronal wave in an inclined eruptions (panel (a)), the lowest part of the wavefront is close to the nose part of the shock wave, hence it is relatively strong at the onset (at $t_0$). The expanding lowest part exerts a strong compression towards the solar surface resulting in the bright segment (the red patch). The compression ignites a pressure jump that will cause a density/temperature increase not only in the TR but also in the chromosphere, in the form of wave signatures in the TR and chromosphere (the distortions and the projected arcs). When the nose part of the coronal wavefront lifts higher later (at $t_1$), the lowest part, being further away from the nose part of the coronal wave, becomes so weak that it hardly compresses the TR beneath, and consequently the wave signatures in the TR and chromosphere vanish. On the contrary, in the radial eruption scenario (panel (b)), the nose of coronal wave initially forms at a typical height of 50--100 Mm~\citep{Patsourakos2012} a few minutes after the early rising phase of the eruption core. The lowest rim of the coronal wavefront is far from the nose part of the shock (above a height of $h$) and becomes too weak to imprint the TR and the chromosphere. 

\begin{figure}[!ht]
\centering
\includegraphics{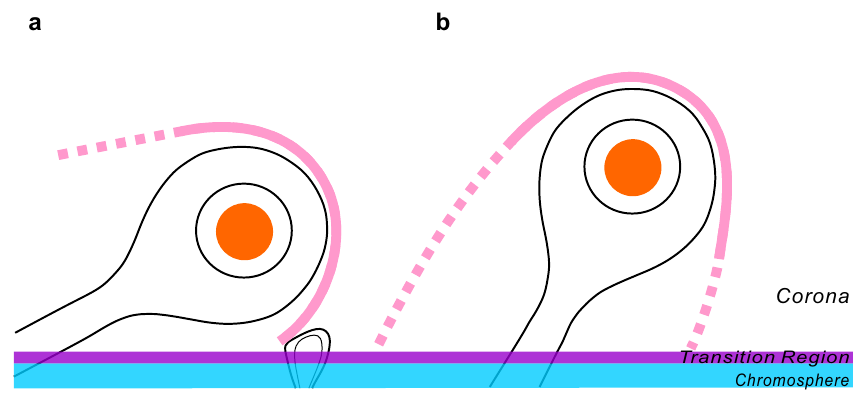}
\caption{{\bf The unsuccessful situations for Moreton waves in inclined eruptions.} {\bf a)}: The coronal wave encounters the low-lying coronal loops. {\bf b)}: The inclination angle for the eruption is large. The coronal wave (the solid-dotted pink curve) consists of a strong nose (solid) and weak flanks (dotted).
}
\label{f5}
\end{figure}

In this scenario, Moreton waves are easily generated by coronal waves that are associated with inclined eruptions. However, would all coronal waves with inclined eruptions leave imprints in the chromosphere? We show two more Moreton waves associated with inclined eruptions (Supplementary Fig.~\ref{fs9} and Fig.~\ref{fs9} animation). The Moreton wave associated with an X1.8 flare on 2011 September 7 was hardly distinguishable, and was only confirmed by the oscillation of a low-lying filament and nearby loops (the pink and black arrows in top panels in Fig.~\ref{fs9}). The Moreton wave (blue arrows) associated with an X1.0 flare on 2021 October 28 was also weak, and the westward erupting prominence (yellow arrows) had an inclination angle of $\sim$30$^{\circ}$ calculated with reconstructed points of the erupting prominence in dual perspectives (bottom panels in Fig.~\ref{fs9}). Hence, the scenario can be complimented by the factors of the scale of magnetic structures beneath the inclined coronal wavefronts and the range of inclination angles. Assuming that there exists an underlying large-scale coronal loop system that would be intact during the downward compression of the coronal wavefronts (Fig.~\ref{f5}a), the large-scale loop system possibly invokes the MHD wave refractions from regions of high-Alv{\'e}n speed \citep{Wang2000}. The refraction from the ambient large-scale loop systems indeed contributed to the lack of Moreton waves for some coronal waves in inclined eruptions. On the other hand, if the inclination angle of the eruption is small to the radial direction, it should become difficult for the coronal wave to disturb the TR and the chromosphere significantly, and the Moreton wave signature will be faint or may even disappear (Fig.~\ref{f5}b). 

Therefore, the scenario introduced above suggests that the reliance on the highly inclined eruptions is the key to resolving the rarity puzzle of Moreton waves. Hence, the comprehensive relationship between the coronal waves and Moreton waves is likely established by the highly inclined eruptions, with the additional factors (the inclined angle and the magnetic structures beneath downward coronal wavefronts) for the scenario. We also imply that the wave in He II 304~{\AA} may be the counterparts of Moreton waves when the Moreton waves are undetectable in low-quality H$\alpha$ data. According to this scenario, a coronal wave related to an impulsive X-class flare need a moderately inclined eruption, and a coronal wave associated with a weak B-class flare can ignite a detectable Moreton wave when the eruption and the associated coronal wavefront are extremely ``inclined". The high-quality chromospheric data from the Chinese H$\alpha$ Solar Explorer~\citep{Li2022}, launched in October of 2021, will provide more cases of Moreton waves to check the scenario. Further simulations are demanded to verify the scenario proposed in this Letter.

\begin{acknowledgments}
SDO is a mission of NASA's Living With a Star Program. We gratefully acknowledge the usage of data from the SDO, STEREO, and from the ground-based GONG project. This work is supported by grants of NSFC 11790303, 12073016, 11825301, 11973031, and 12127901. R.E. is also grateful to the Science and Technology Facilities Council (STFC, grant No. ST/M000826/1) for the support received.
\end{acknowledgments}
\newpage

\newpage

\newpage

\newpage
\newpage

\newpage

%





\newpage
\appendix

\section{Appendix information}
\renewcommand{\thefigure}{S\arabic{figure}}

\bigskip{}
\noindent{\bf Data}
The analysis of coronal waves is primarily based on the extreme ultraviolet (EUV) observations in dual perspectives from AIA 304, 171, 193~{\AA} and EUVI 304, 171, and 195~{\AA} (Fig.~\ref{f1}, Fig.~\ref{f2}, and associated animations). Each AIA image has a cadence of 12 second and a pixel size of 0.6$''$. EUVI images have a pixel size of 1.58$''$, and their cadences are 2.5 minutes, 5 minutes or 10 minutes. Moreover, to study the evolution of Moreton waves, we use the H$\alpha$ filtergrams from the Global Oscillation Network Group (GONG) of the National Solar Observatory. The H$\alpha$ images have a cadence of 1 minute and a pixel size of $\sim$1$''$ (Fig.~\ref{f2}).

\begin{figure}[!ht]
\centering
\includegraphics{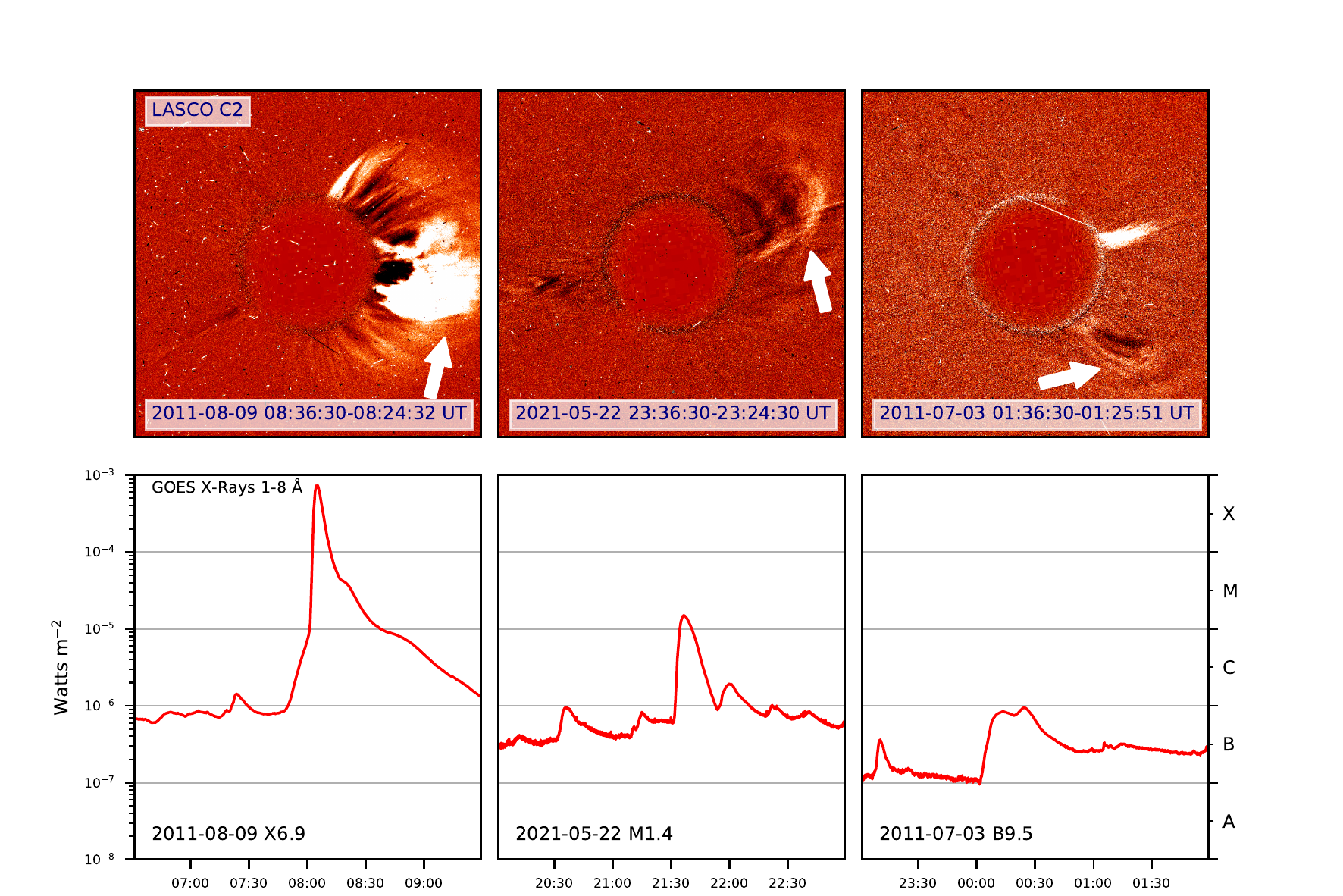}
\caption{{\bf The associated CMEs and flares.} Upper row: The related CMEs (white arrows) are shown in difference images of LACSO C2. Bottom row: The related flares are indicated by GOES X-rays 1--8~{\AA}.
}
\label{fs1}
\end{figure}

\begin{figure}[!ht]
\centering
\includegraphics{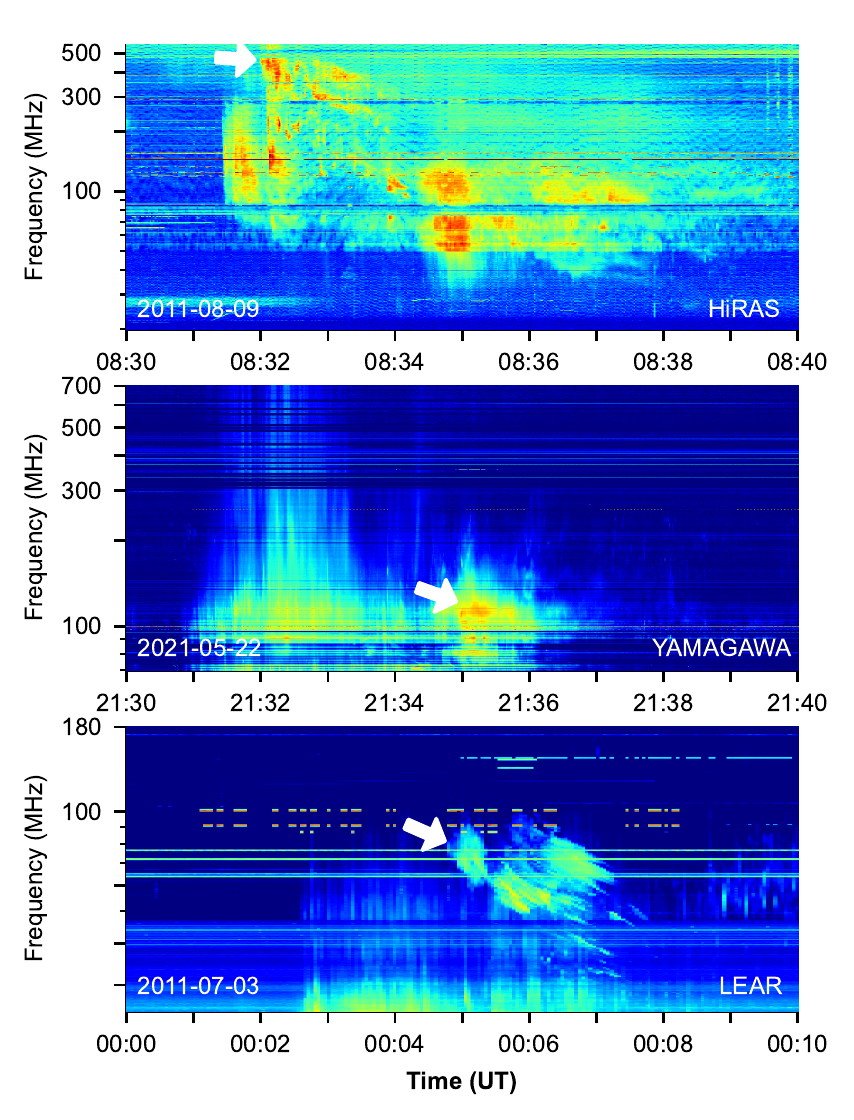}
\caption{{\bf The associated radio bursts.} The radio dynamic spectrums for three cases are obtained from HiRAS, YAMAGAWA, and LEAR, respectively, and the white arrows indicate the onsets of radio bursts.
}
\label{fs2}
\end{figure}

The evolution of the CMEs in the high corona (white arrows) is shown by Large Angle and Spectrometric Coronagraph (LASCO)~\citep{Brueckner1995} onboard the {\it Solar and Heliospheric Observatory (SOHO)} spacecraft, and the related flares are recorded by the Geostationary Operational Environment Satellite (GOES) in the form of integrated full-disk soft X-ray emission from the Sun (Supplementary Fig.~\ref{fs1}). In addition, the associated radio bursts (white arrows) was detected in the frequency range of 25-500 MHz by the metric spectrometers from Hiraso RAdio Spectrograph (HiRAS) and YAMAGAWA of National Institute of Information and Communications Technology (NICT)~\citep{Kondo1995, Iwai2017}, and Learmonth (LEAR) of the radio Solar Telescope Network (RSTN)~\citep{Kennewell2003} (Supplementary Fig.~\ref{fs2}).

\begin{figure}[!ht]
\centering
\includegraphics{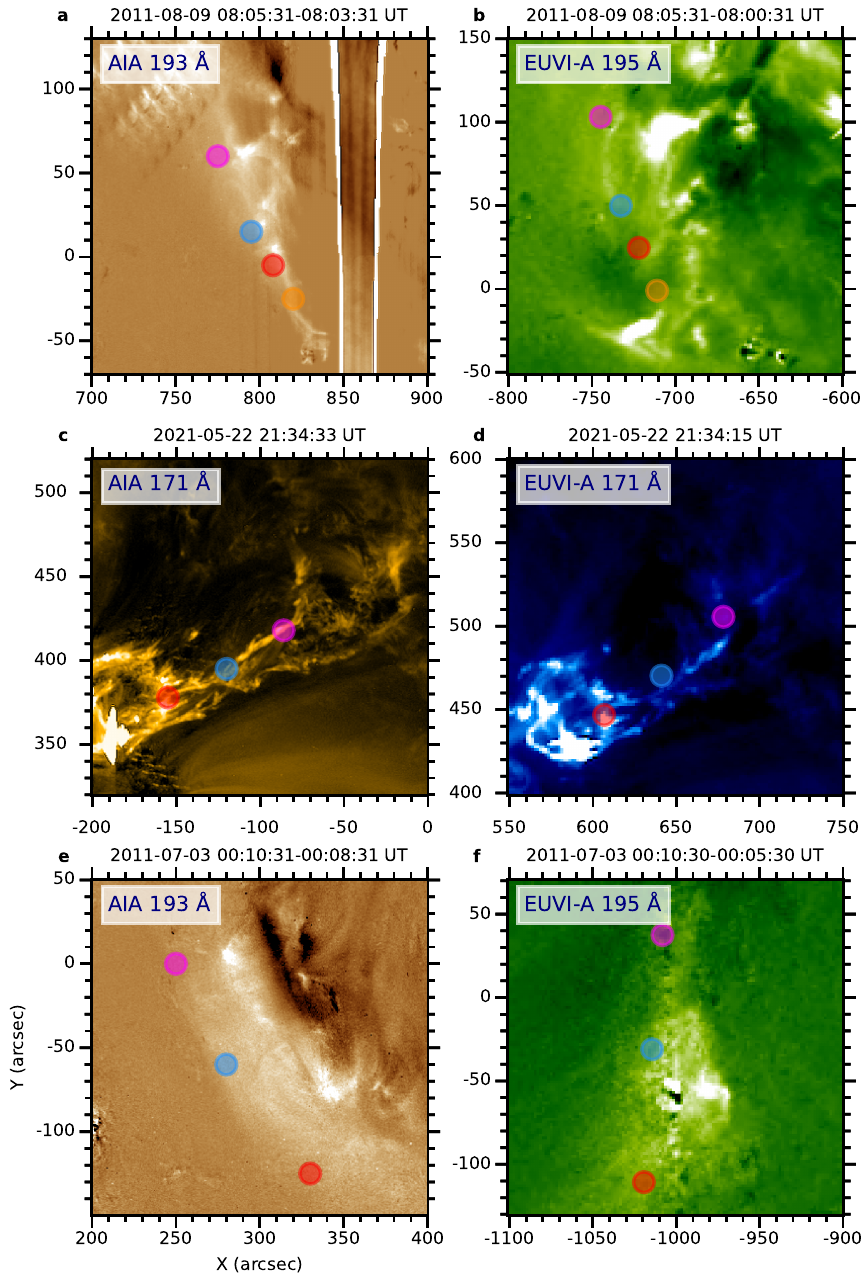}
\caption{{\bf The selected points for the reconstructions.} The colored points in dual perspectives from AIA and EUVI-A represent the positions for the reconstructions in Supplementary Fig.~\ref{fs4}.
}
\label{fs3}
\end{figure}

\begin{figure}[!ht]
\centering
\includegraphics{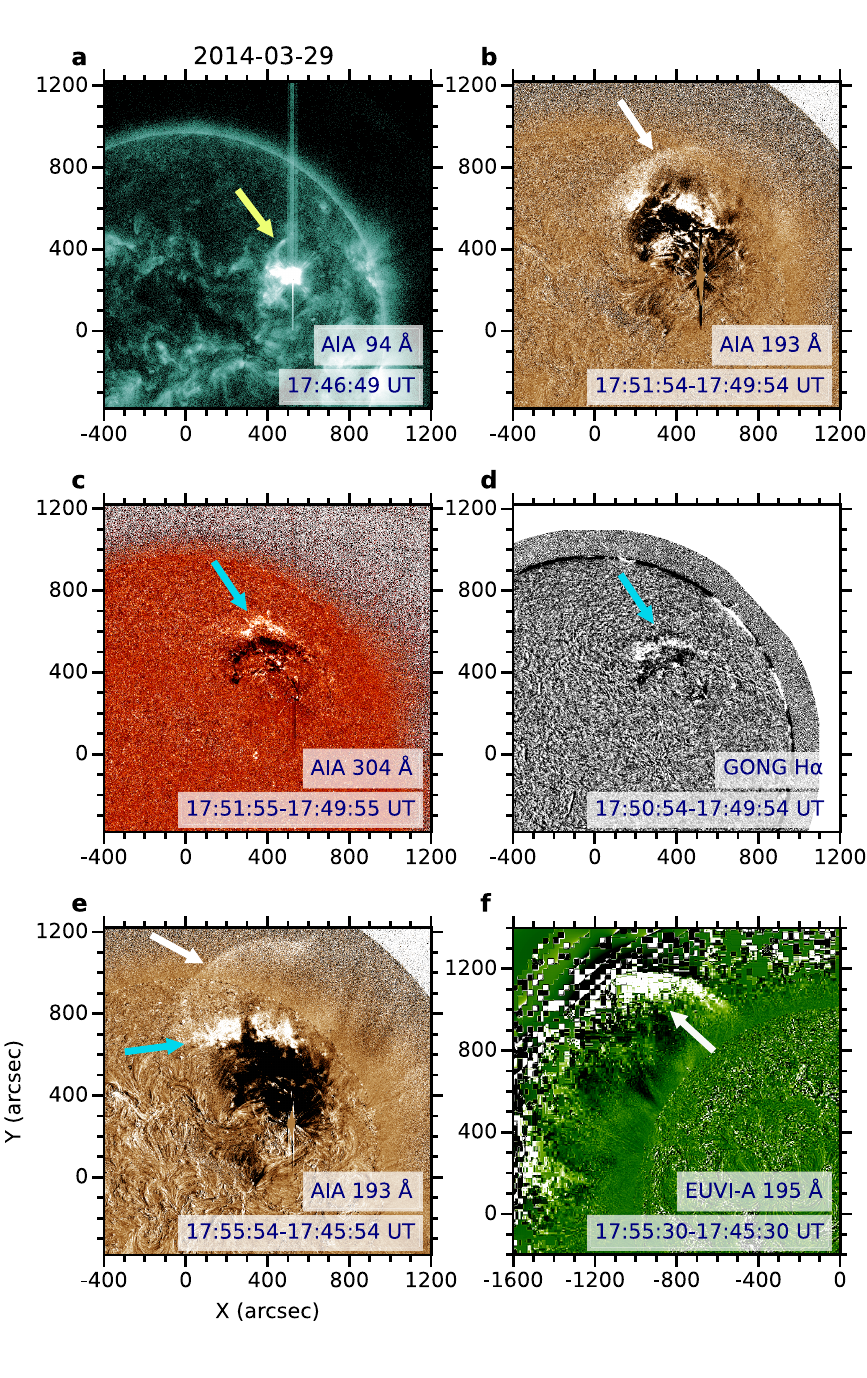}
\caption{{\bf The Moreton wave with an X1.0 flare on 2014 March 29.} The inclined erupting structure (the yellow arrow) is shown in AIA 94~{AA}. In the difference images in dual perspectives, the blue arrows show the bright segments of coronal waves (white arrows) in AIA 193~{\AA} and EUVI 195~{\AA} and the simultaneous wave signatures in AIA 304~{\AA} and GONG H$\alpha$.
}
\label{fs6}
\end{figure}

\begin{figure}[!ht]
\centering
\includegraphics{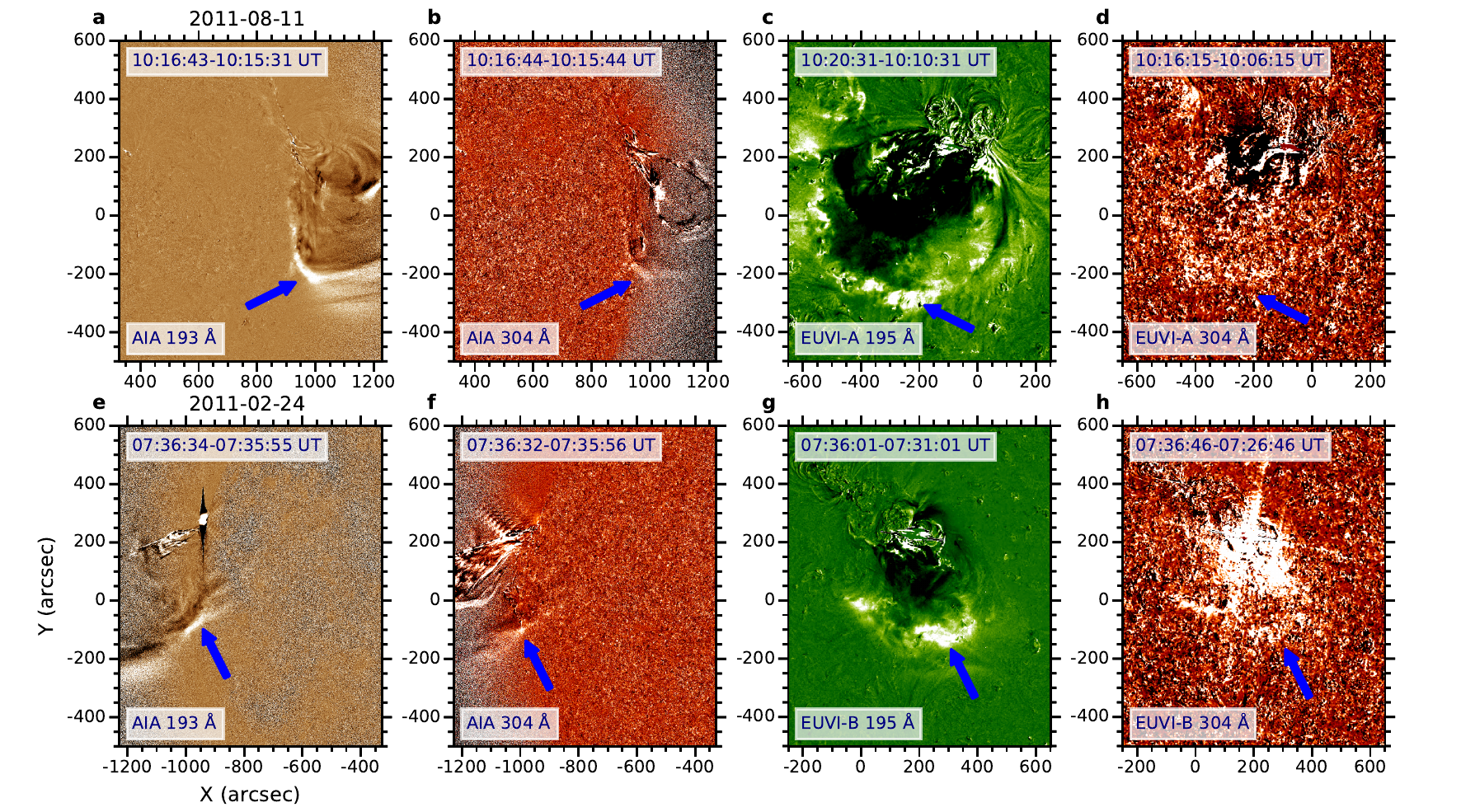}
\caption{{\bf Two cases of inclined limb eruptions.} Two eruptions are separately accompanied with a C6.2 flare on 2011 August 11 and an M3.5 flare on 2011 February 24. In the difference images in dual perspectives, the blue arrows show the bright segments of coronal waves in AIA 193~{\AA} and EUVI 195~{\AA} and the simultaneous wave signatures in He II 304~{\AA}. (An animation of this figure is available online.)
}
\label{fs7}
\end{figure}

\bigskip{}
\noindent{\bf Methods}
To identify well the wavefronts and CMEs, the image is subtracted by one previous image, and the times of two images are attached (Fig.~\ref{f1}, Fig.~\ref{f2}, Supplementary Fig.~\ref{fs1}, Fig.~\ref{fs3}, Fig.~\ref{fs6}, Fig.~\ref{fs7}, and associated animations). The wave signatures in different layers are analysed by the time-slice approach, in which a time-distance plot was constructed with a stack of slices along a selected sector for a set of difference images. The wave speeds and errors (blue lines) are obtained by fitting with the linear ({\it linfit.pro}) function in the SolarSoftWare (SSW), assuming a measurement uncertainty of 4 pixels ($\sim1.74$ Mm) for the selected points (Supplementary Fig.~\ref{fs5}).

Benefiting with the observations in dual perspectives (upper panels in Supplementary Fig.~\ref{fs4}), we derive the real locations of the erupting cores (the core material of erupting filaments/prominences and jets) and wavefronts by the reconstruction procedure of {\it scc\_measure.pro} in SSW. To estimate the eruption inclination, we reconstruct some selected points from the erupting core of E2 both that simultaneously appear in AIA 171~{\AA} and EUVI-A 171~{\AA}. For E1 and E3, we choose the bright segment instead of invisible erupting cores (Supplementary Fig.~\ref{fs3}), assuming that the inclination of the erupting core is approximately equal to that of the bright segment (the orange oval core and the red segment in Fig.~\ref{f4}). Finally, we superimpose the reconstructed points on one AIA 304~{\AA} image in which the eruption center is rotated to the limb (bottom panels in Supplementary Fig.~\ref{fs4}).

We also investigated the emission evolution in the fronts of coronal waves with the differential emission measure (DEM) method by employing the sparse inversion code~\citep{Cheung2015, Su2018}. The emission at a range of temperature is obtained by a set of AIA images in six channels, i.e., 131~{\AA} (Fe XXI, $\sim$10 MK), 94~{\AA} (Fe XVIII, $\sim$6.4 MK), 335~{\AA} (Fe XVI, $\sim$2.5 MK), 211~{\AA} (Fe XIV, $\sim$2.0 MK), 193~{\AA} (Fe XII, $\sim$1.6 MK), and 171~{\AA} (Fe IX, $\sim$0.6 MK). The DEM evolution plots are constructed with the average value in the selected regions (cyan boxes in Fig.~\ref{f2}) for a series of EM maps (bottom panels in Fig.~\ref{f3}).










\newpage

\newpage

\newpage

\newpage
\begin{figure}[!ht]
\centering
\includegraphics[width=0.95\linewidth]{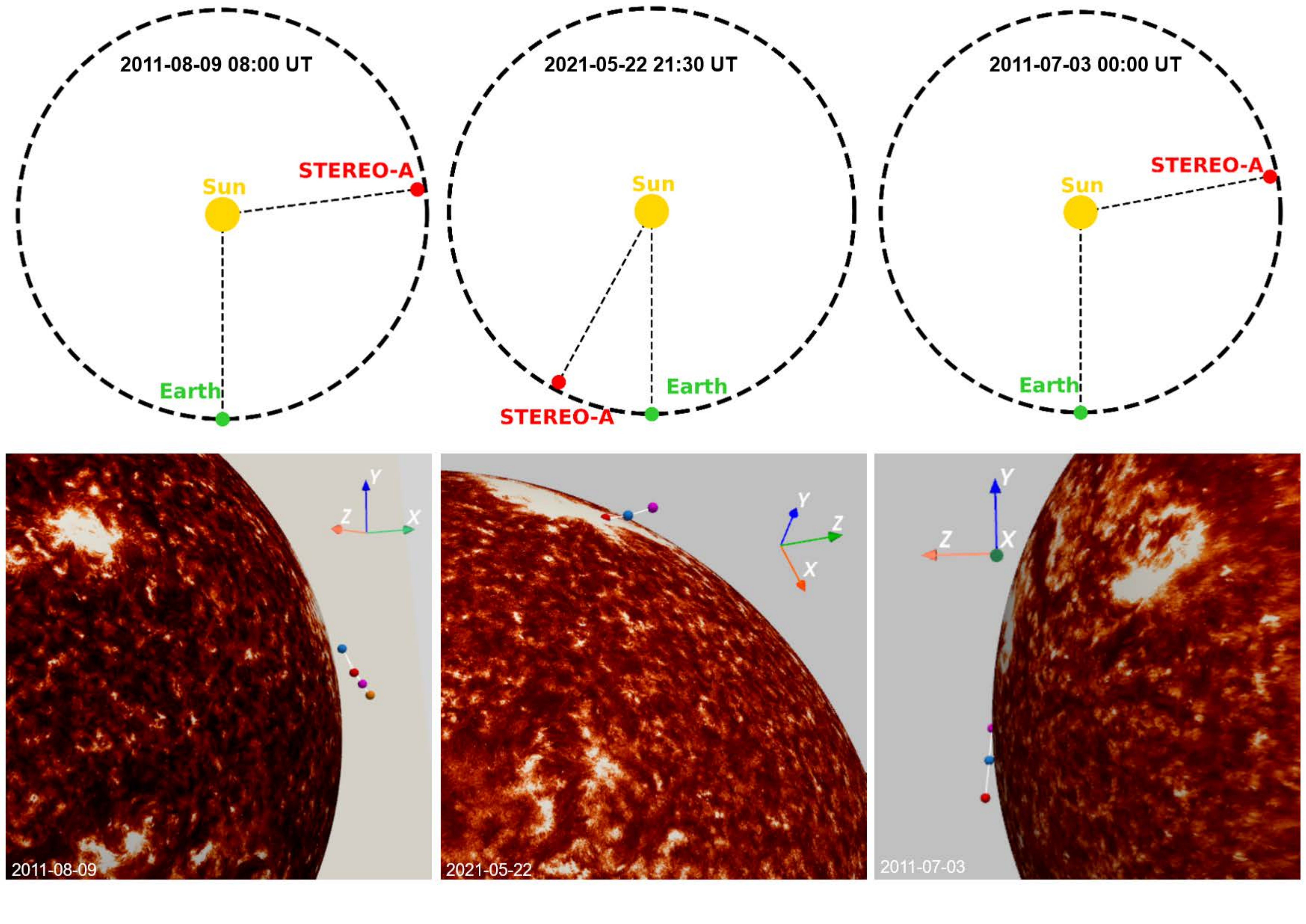}
\caption{{\bf The inclination angles of the eruptions.} Upper row: The positions of the Sun, the Earth, and the STEREO-A. Bottom row: The selected points in the limb of rotated AIA 304~{\AA} images to show the the inclination angles that are $\sim73^{\circ}$, $\sim63^{\circ}$, and $\sim76^{\circ}$, respectively. The axis of $z$ points to the Earth, and the axe of $x$ and $y$ define the coordinate plane of AIA images.
}
\label{fs4}
\end{figure}

\newpage
\begin{figure}[!ht]
\centering
\includegraphics{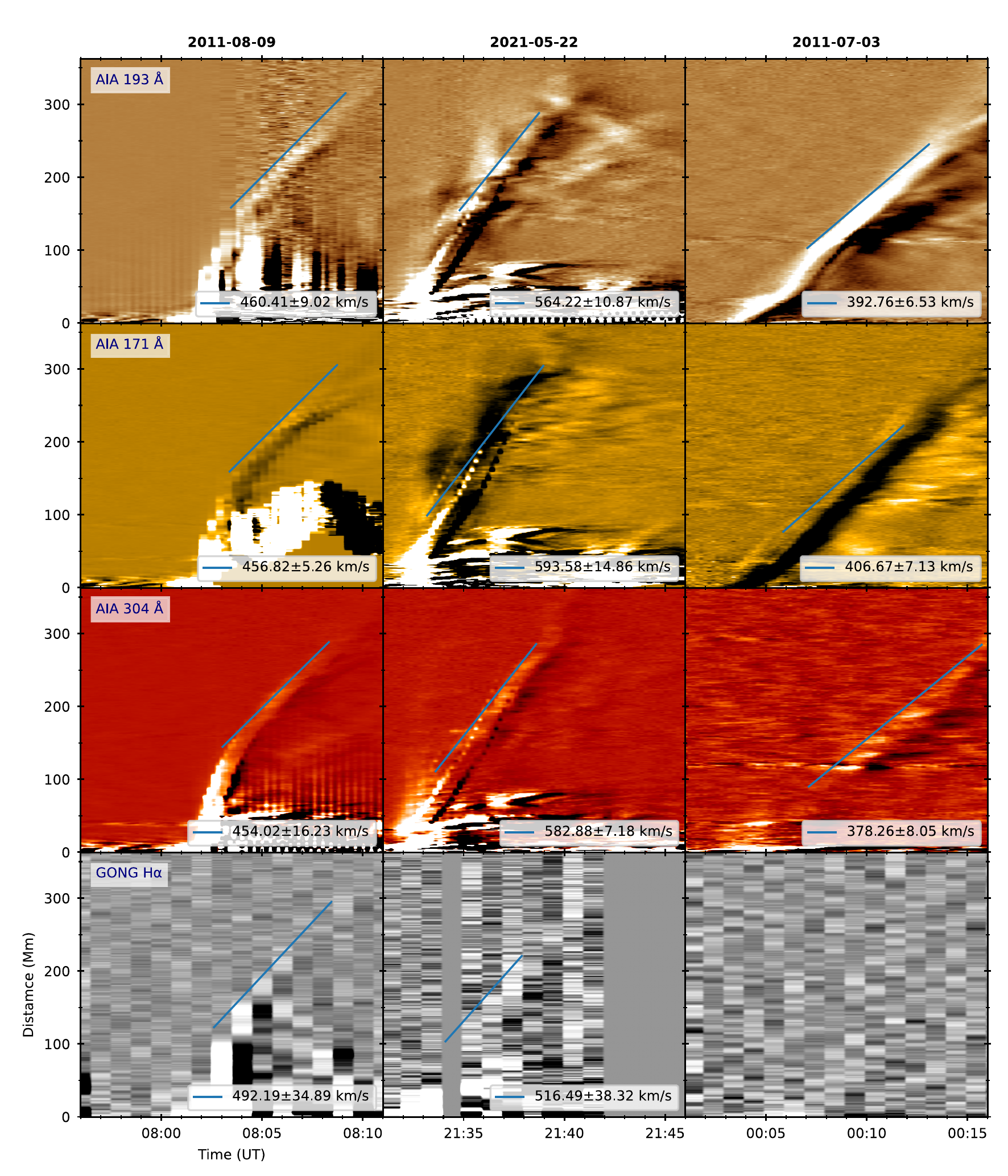}
\caption{{\bf Wave velocities in different layers.} The velocities in AIA 193, 171, 304~{\AA} and GONG H$\alpha$ are calculated along the sectors in Fig.~\ref{f1} for three cases (left, middle, right columns), and the inclined lines indicate the measured speeds. Note that the chromospheric wave signature for E3 is hardly distinguished in the last panel.
}
\label{fs5}
\end{figure}

\newpage

\newpage

\newpage
\begin{figure}[!ht]
\centering
\includegraphics{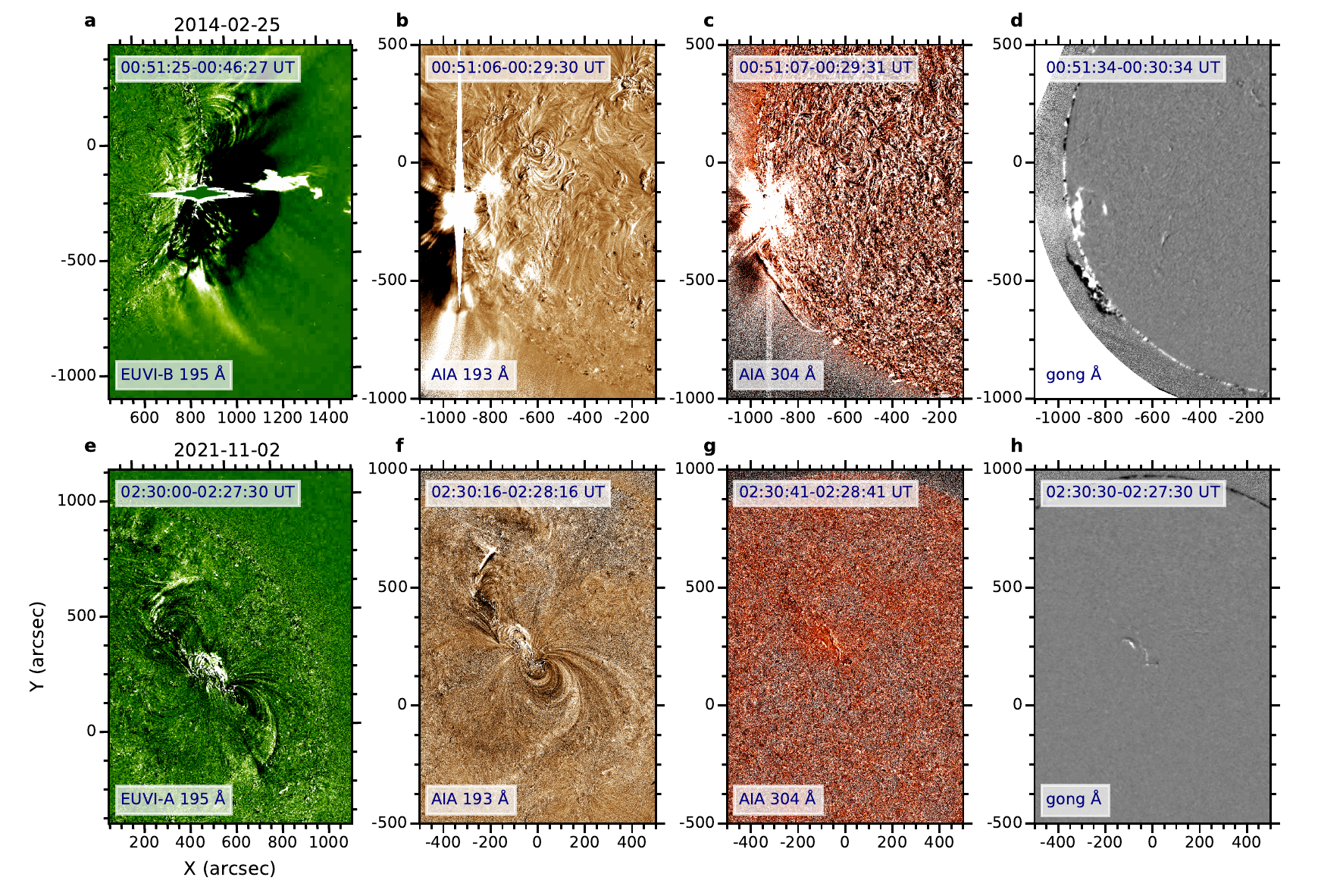}
\caption{{\bf Two cases of radial eruptions.} Two eruptions are separately accompanied with an X4.9 flare on 2014 February 25 and an M1.7 flare on 2021 November 2. In the difference images in dual perspectives, the coronal waves are obvious in AIA 193~{\AA} and EUVI 195~{\AA}, but there is no wave signature in He II 304~{\AA} and H$\alpha$. (An animation of this figure is available online.)
}
\label{fs8}
\end{figure}

\newpage
\begin{figure}[!ht]
\centering
\includegraphics{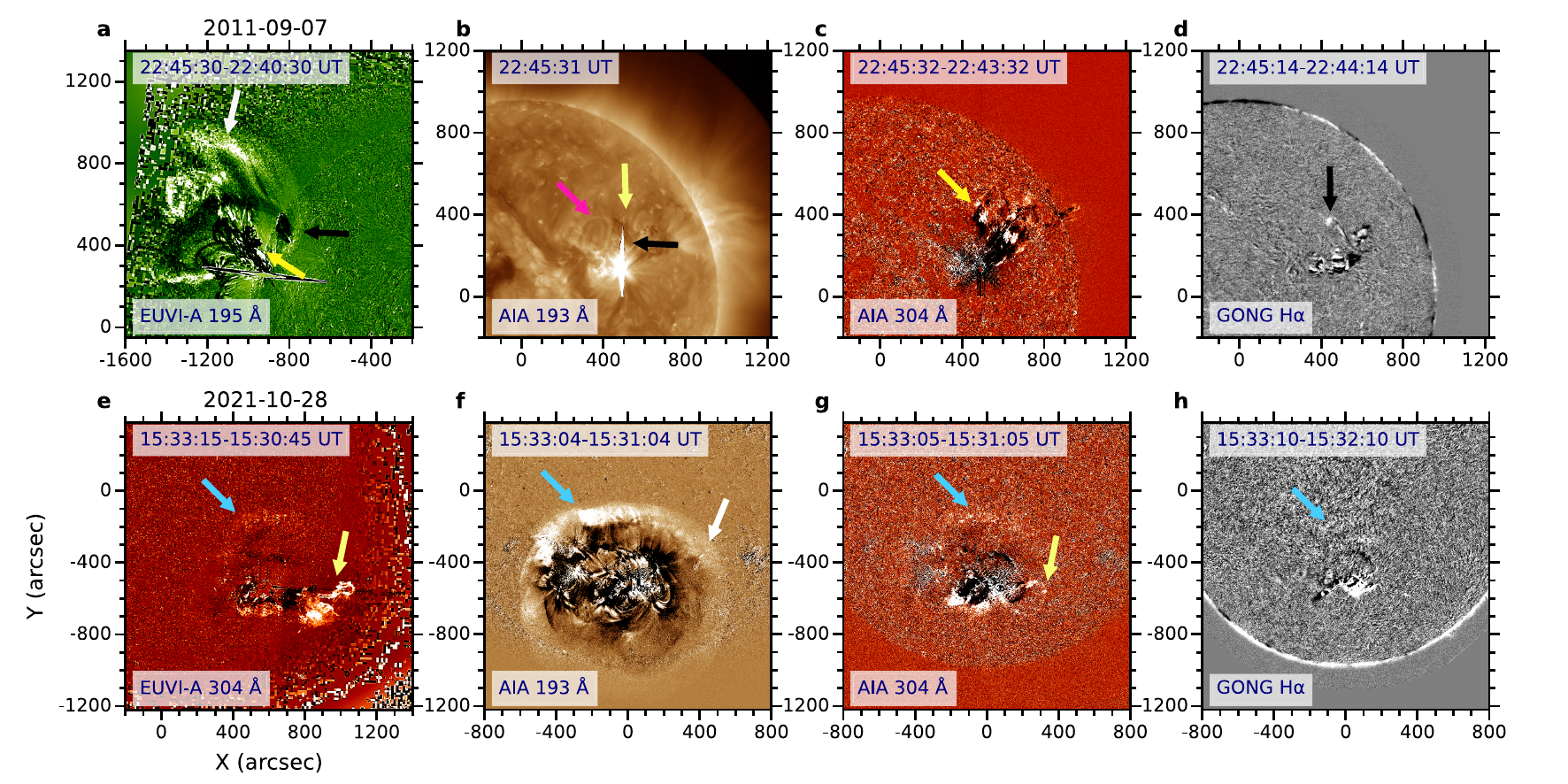}
\caption{{\bf Two more cases of inclined eruptions.} Two eruptions are separately accompanied with an X1.8 flare on 2011 September 7 and an X1.0 flare on 2021 October 28. In the difference images in dual perspectives, the blue arrows show the bright segments of coronal waves (white arrows) in AIA 193~{\AA} and EUVI 195~{\AA} and the simultaneous wave signatures in AIA 304~{\AA} and GONG H$\alpha$. The waves are closely associated with inclined erupting filaments (yellow arrows), and are affected by the ambient coronal loops (the pink arrow) and the underlying filament (black arrows). (An animation of this figure is available online.)
}
\label{fs9}
\end{figure}
\section*{}
\bibliography{sample631}{}
\bibliographystyle{aasjournal}


\end{CJK*}
\end{document}